%% file: arxiv.tex
\documentclass[12pt]{article}
\usepackage{times}
\usepackage{geometry}
\usepackage{enumitem,amssymb}
\usepackage{ragged2e}

\input{definitions_2017.tex}

\usepackage{microtype}

\usepackage[utf8]{inputenc}
\usepackage[noblocks,affil-it]{authblk}

\usepackage{wrapfig}
\usepackage{natbib}
\usepackage{amsmath,booktabs}
\usepackage{graphicx}

\usepackage{txfonts}
\usepackage[breaklinks,colorlinks=true,linkcolor=blue,citecolor=blue, urlcolor=blue]{hyperref}

\usepackage{mathptmx}
\usepackage{indentfirst}
\usepackage{sectsty, titlesec}
\sectionfont{\large}

\newlist{thematic}{itemize}{8}
\setlist[thematic]{label=$\square$}
\usepackage{pifont}
\newcommand{\cmark}{\ding{51}}%

\newcommand{\done}{\rlap{$\square$}{\raisebox{2pt}{\large\hspace{1pt}\cmark}}%
\hspace{-2.5pt}}

\begin{document}

\begin{raggedright}



{\huge Astro2020 Science White Paper} \linebreak \linebreak
{\huge Time Domain Studies of Neutron Star and Black Hole Populations: X-ray Identification of Compact Object Types \linebreak \par} 

\normalsize

\noindent \textbf{Thematic Areas:} \hspace*{55pt} $\square$ Planetary Systems \hspace*{5pt} $\square$ Star and Planet Formation \hspace*{20pt}\linebreak
$\done$ Formation and Evolution of Compact Objects \hspace*{24pt} $\square$ Cosmology and Fundamental Physics \linebreak
$\square$  Stars and Stellar Evolution \hspace*{1pt} $\square$ Resolved Stellar Populations and their Environments \hspace*{40pt} \linebreak
  $\square$    Galaxy Evolution   \hspace*{49pt} $\square$ Multi-Messenger Astronomy and Astrophysics \hspace*{65pt} \linebreak
  
\textbf{Principal Author:}

Name: Neven Vulic$^{1,2,3}$
 \linebreak						
Institution: (1) NASA Goddard Space Flight Center (2) Center for Research in Space Sciences and Technology II (3) University of Maryland, College Park
 \linebreak
Email: neven.vulic@nasa.gov
 \linebreak
Phone: +1 (301) 286-8672
 \linebreak
 
\textbf{Co-authors:} A. E. Hornschemeier (NASA GSFC/John Hopkins University), V. Antoniou (Texas Tech University/Harvard CfA), A. R. Basu-Zych (NASA GSFC/CRESSTII/University of Maryland Baltimore County), B. Binder (California State Polytechnic University), F. M. Fornasini (Harvard CfA), F. Furst (ESAC), F. Haberl (MPE), M. Heida (Caltech), B. D. Lehmer (University of Arkansas), T. J. Maccarone (Texas Tech University), A. F. Ptak (NASA GSFC/John Hopkins University), G. R. Sivakoff (University of Alberta), P. Tzanavaris (University of Maryland Baltimore County/NASA GSFC), D. R. Wik (University of Utah), B. F. Williams (University of Washington), J. Wilms (Universitat Erlangen-Nurnberg), M. Yukita (John Hopkins University), A. Zezas (University of Crete/Harvard CfA)
  \linebreak

\end{raggedright}

\section*{Abstract}

What are the most important conditions and processes governing the growth of stellar-origin compact objects?
The identification of compact object type as either black hole (BH) or neutron star (NS) is fundamental to understanding their formation and evolution. To date, time-domain determination of compact object type remains a relatively untapped tool.  Measurement of orbital periods, pulsations, and bursts will lead to a revolution in the study of the demographics of NS and BH populations, linking source phenomena to accretion and galaxy parameters (e.g., star formation, metallicity).  To perform these measurements over sufficient parameter space, a combination of a wide-field ($>5000$ deg$^{2}$) transient X-ray monitor over a dynamic energy range $(\sim1-100$ keV) and an X-ray telescope for deep surveys with  $\lesssim5$\arcsec\ PSF half-energy width (HEW) angular resolution are required.  Synergy with multiwavelength data for characterizing the underlying stellar population  will transform our understanding of the time domain properties of transient sources, helping to explain details of supernova explosions and gravitational wave event rates.

\pagebreak

\section{Introduction}

The end result of the evolution of massive stars is the production of a population of NS and BH. These stellar-origin compact objects grow through two main channels: accretion and mergers. The accretion growth, which is likely the dominant mode, is affected by star formation, stellar evolution, and binary processes such as the strength of stellar winds, the common envelope phase, and the initial mass function. Merger growth (e.g., gravitational wave sources) gives us a unique snapshot view of the masses of stellar origin compact objects, which is also connected to stellar evolution and star formation processes in the Universe. It is critically important that we better understand NS/BH populations. As remnants of supernova explosions and the end states of massive stars, they allow us to probe key phases in stellar evolution and death. Also, in addition to understanding the progenitor paths for gravitational wave sources, the X-ray emission from this population likely plays an important role in the early heating of the primordial intergalactic medium (IGM) at $10 < z < 20$ \citep[e.g.][]{fragos10-13, mesinger04-14, pacucci09-14, madau05-17, sazonov06-17, das07-17}.

We have had an explosion of information on the overall energy output (including refinement of key population properties) from accreting compact objects detected via X-ray emission over the age of the Universe, thanks to a suite of X-ray observatories \citep[e.g.][]{lehmer11-10, stiele10-11, mineo01-12, fragos10-13, mineo01-14, lehmer07-14, basu-zych02-16, haberl02-16, peacock02-16, vulic09-18}. Key breakthroughs have come via X-ray imaging by \chandra/\xmmn\ in the $0.3-10$ keV energy band and \nustar\ at $E>10$ keV, combined with emerging multi-messenger constraints on mergers from gravitational wave observatories such as LIGO. Properties of NS and BH populations have been reliably connected to galaxy properties such as star formation rate (SFR), stellar mass (M$_{\star}$), stellar age, and metallicity, and has been studied over significant intervals of cosmic time. 
However, nearby galaxy surveys have long classified X-ray binaries (XRBs) by the mass category of their donor stars, high-mass (HMXB) and low-mass (LMXB). The identification of the compact-object type has been limited to XRBs in our Galaxy, the Magellanic Clouds, and a few of the brightest nearby extragalactic systems.

There are a number of methods that have been used to determine the compact object type in XRBs. The most reliable to date are in the X-ray time domain.
The detection of coherent pulsations and/or a Type I X-ray burst provides confirmation that the object is a NS. Otherwise, Kepler's third law is required to determine the mass of the compact object in the XRB and classify it as a NS or BH. Quasiperiodic oscillations from the ultraluminous X-ray sources (ULX) M82 X-1 and NGC 1313 X-1 \citep{pasham09-14, pasham09-15} have also been used to obtain mass estimates for candidate BHs.
These methods require high signal-to-noise data only available for a handful of the brightest extragalactic sources. Time domain studies of compact object {\it populations} have thus not yet been possible outside the Milky Way and Magellanic Clouds.

\textbf{Compact object formation is important for a wide range of topics in astronomy and astrophysics.}  We can improve our understanding of hard-to-model supernova explosions if we more accurately map them to their resulting stellar remnants. The demographics of BH and NS populations are required to determine formation rates for gravitational wave events. Here we discuss some important, yet poorly understood populations (e.g., ULXs, pulsars, Wolf-Rayet XRBs, and some enigmatic ultraluminous burst sources) that would yield key insight into these topics via X-ray measurements of spin/orbital periods, bursts, and complementary multiwavelength constraints.

\section{Ultraluminous X-ray Sources \& Pulsar Populations} \label{sec:ulx}

\begin{wrapfigure}{R}{0.5\textwidth}		
   \vspace{-0.75cm}
  \begin{center}
    \includegraphics[width=0.48\textwidth]{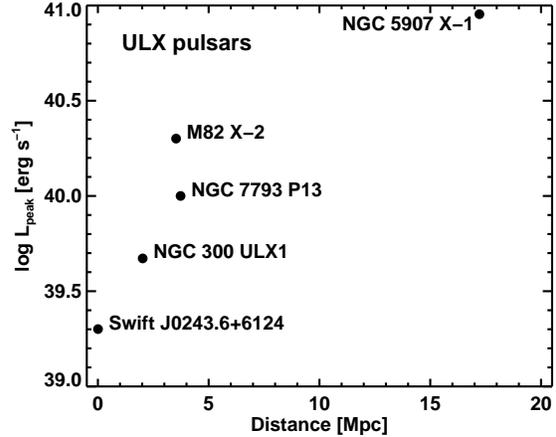}
  \end{center}
     \vspace{-0.75cm}
\caption{\small{{\bf Some NSs emit as much as 100 times above the Eddington limit and we do not yet understand how.} Peak \lx\ ($0.5-10.0$ keV) vs.\ host galaxy distance for currently known ULX pulsars. The sensitivity and collecting area of current X-ray telescopes limit the detection of ULX \textit{pulsars} having \lx\ $\simeq10^{39}$ \es\ to $d \approx5$ Mpc.}}
  \vspace{1.5cm}
  \label{fig:ulxpul}
\end{wrapfigure}

ULXs are off-nuclear X-ray sources having X-ray luminosities \lx$\gtrsim10^{39}$ \es\ ($0.5-10$ keV) that exceed the Eddington luminosity for a $\approx10$ \msun\ BH, assuming isotropy.  ULXs were generally thought to be massive, possibly-beamed,  stellar-mass BHs, or intermediate-mass BHs accreting at sub-Eddington rates.
The detection of coherent pulsations from 4 ULXs in nearby galaxies\footnote{Also the candidate ULX pulsar M51 ULX8, inferred from the cyclotron resonance scattering feature \citep{brightman04-18}.} with \xmmn/\nustar\ and one source in the Milky Way (detected with \swift\ and \fermi) has demonstrated that NSs are capable of producing the high luminosities that previously were thought to be the domain of BH systems \citep{bachetti10-14, furst11-16, furst01-17, israel02-17, israel03-17, carpano05-18, kosec09-18, walton04-18, wilson-hodge08-18}.
Figure \ref{fig:ulxpul} shows the peak \lx\ of known ULX pulsars as a function of distance.

\citet{wiktorowicz09-17} simulated isolated binaries using the {\sc startrack} population synthesis code, finding that NS were the dominant ULX accretors a few hundred Myr post-starburst.
Current X-ray telescopes have the ability to detect pulsations from ULXs with \lx\ $\simeq10^{39}$ \es\ out to $d \approx5$ Mpc ($\sim20\%$ pulsed fraction). 
We are therefore unable to search for pulsations within the known population of $>400$ ULXs (all with \lx\ $\gtrsim10^{39}$ \es) that have been detected out to $d \approx200$ Mpc, even though 80\% are within 50 Mpc \citep{liu01-11, walton09-11, earnshaw03-19}. 
Detailed surveys of nearby ULXs are necessary for statistically-significant studies of their population characteristics (e.g., variability/pulsed fraction, age dependence, spin period distribution) in addition to compact object identification.

X-ray timing analysis can also be extended to the general (non-ULX) X-ray pulsar population in a range of formation environments in nearby galaxies. Currently, X-ray pulsar \textit{populations} can only be studied in the Milky Way and Magellanic Clouds \citep[e.g.][]{haberl02-16, antoniou06-10, antoniou06-16}. In a single year, the \swift\ SMC survey detected Type I X-ray bursts and orbital periods for 6 \be-XRBs \citep{kennea11-18}, systems that are useful for predicting gravitational wave event rates. To extend ULX pulsar detection from $d \approx5$ to 25 Mpc and study non-ULX pulsars beyond the Magellanic Clouds, an X-ray telescope must have sufficient throughput to detect sources with flux $\sim10^{-17}$ erg cm$^{-2}$ s$^{-1}$ within 200~ks, as well as a large field of view ($\sim0.5$ deg$^{2}$) for efficient surveys of galaxies with large angular sizes (e.g., M31 and the Magellanic Clouds), and a $\lesssim5\arcsec$ PSF HEW to avoid confusion in star-forming regions.

\section{Wolf-Rayet XRBs \& Massive Stellar-Mass BH Production}	

\begin{wrapfigure}{r}{0.5\textwidth}		
   \vspace{-0.75cm}
  \begin{center}
    \includegraphics[width=0.48\textwidth]{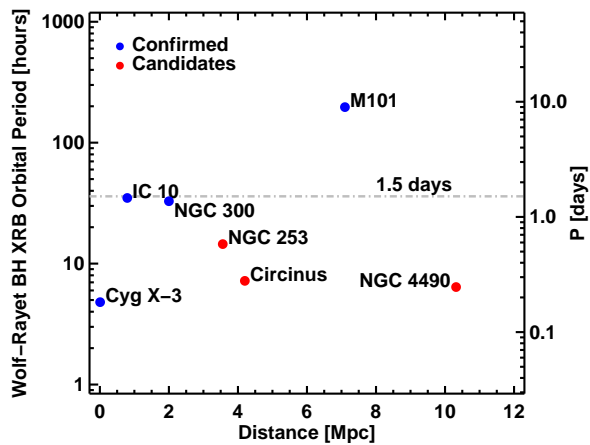}
  \end{center}
     \vspace{-0.75cm}
\caption{\small{The confirmed (blue) and candidate (red) Wolf-Rayet BH-XRBs in the Universe. Shown is their orbital period vs.\ distance of the host galaxy. WR XRBs below the horizontal line at $\sim1.5$ days are expected to form a BH-BH binary and merge within a Hubble time.}}
  \label{fig:wrxrb}
\end{wrapfigure}

LIGO/Virgo have now discovered seven binary BH mergers with pre-merger (individual BH) masses $\simgt 30$ \msun\ \citep{the-ligo-scientific-collaboration11-182}. These are comparable to the most massive known\footnote{BH masses for IC 10 X-1 and NGC 300 X-1 were likely overestimated \citep{laycock01-15, laycock09-15, binder08-15}.} stellar-mass BHs Cyg X-1 and M33 X-7 ($\sim15$ \msun, \citealt{orosz10-07, orosz12-11}). When looking for gravitational wave progenitor populations, massive BH XRBs with orbital periods less than $\sim1.5$ days are excellent candidates, as they will merge within a Hubble time \citep{van-den-heuvel11-17}. Likely candidates for this scenario include Wolf-Rayet (WR) XRBs, which constitute a subclass of HMXBs that have a WR star as their donor, for which, to date, there are only four confirmed examples, namely Galactic source Cyg X-3 \citep{van-kerkwijk10-96} and three additional extragalactic candidates \citep[e.g.][]{esposito09-15}. Fig. \ref{fig:wrxrb} shows WR XRB orbital periods compared to host galaxy distance.

Since WR stars will likely end their lives as BHs, a census of BHs in WR XRB systems is critical to understanding how the Universe is able to produce massive BH mergers. This census can be accomplished by identifying and confirming more of these unique systems in the nearby Universe with next-generation X-ray observations/surveys. \citet{van-den-heuvel11-17} estimated that the Milky Way should have $\sim10$ WR XRBs, about half of which should have luminosities at the Cyg X-3 level (survey completeness throughout the Galactic plane likely hampers detection for other WR XRBs). Due to comparable host environments (star-forming regions) and count rates necessary for detection of ULX pulsations and WR XRB orbital periods, an X-ray telescope with the specifications described in Section \ref{sec:ulx} is required. This would expand the available detection volume by a factor of 100 to $d \approx20$ Mpc, sufficient to create a statistically significant sample of WR XRBs via orbital period measurements. 
Currently, obtaining BH masses is hampered by the difficulty of measuring optical absorption lines to determine the radial velocity amplitude. Future 30-m class telescopes will enable such measurements for many of these systems. Populating the BH mass distribution will put important constraints on supernova explosions by identifying the range of potential remnant masses and gravitational wave events via BH masses in XRBs.

\section{Stochastic Ultraluminous Bursts}

XRBs are known to be highly transient sources, where the phenomenology for state transitions (e.g., quiescence to outburst) in BH and NS systems has been well studied \citep[e.g.][]{maccarone01-03, mcclintock04-06, done12-07, church03-14, tetarenko02-16}. However, outliers have been identified among the extragalactic population that present new challenges to explaining their behavior.

\citet{irwin10-16} recently discovered ultraluminous X-ray bursts in two ultracompact companions of nearby elliptical galaxies in archival \chandra/\xmmn\ data. The flares had rise times of $<1$ min and decay times of $\sim1$ hr, with peak \lx\ of $10^{40-41}$ \es. 
Five other similar flaring sources have been detected by \chandra\
\citep{sivakoff05-05, sun05-13, jonker12-13, glennie07-15,bauer06-17}.
These flares are reminiscent of the mysterious transient extragalactic fast \textit{radio} bursts. 

What leads to seemingly stochastic ultraluminous bursts from X-ray sources? Potential explanations include tidal stripping of a white dwarf onto an intermediate-mass BH, an X-ray afterglow from an off-axis short-duration gamma-ray burst, or a low-luminosity gamma-ray burst at high-redshift \citep{bauer06-17, shen01-19}. However, none of these scenarios can completely explain all the properties of these sources. Identifying the accreting compact object in these burst sources is the first step to understanding the physical mechanisms responsible for this behavior. Any X-ray time domain instrument capable of studying ULXs, pulsars, or WR XRBs will also be able to detect ultraluminous bursts in pointed observations and/or sensitive surveys of nearby galaxies.

\section{Multiwavelength Connections}	

When combined with X-ray observations, multiwavelength data will enhance studies of the various XRB source classes discussed. For ULXs, optical and infrared photometric/spectroscopic observations have been used to confirm ULX distances, counterparts, and probe the surrounding environment \citep[e.g.][]{moon04-11, heida11-15,heida06-16,binder08-18}. For instance, \emph{JWST} will be able to study the obscured star-forming regions surrounding ULXs and Wolf-Rayet XRBs, drawing parallels between local dwarf starburst galaxies and reionization-era analogues\footnote{See white paper by Basu-Zych et al. regarding heating of the early intergalactic medium by HMXBs.}. 
In the optical, the Large Synoptic Survey Telescope (LSST; expected first light in 2021; start of 10-year all-sky survey in 2022) will detect $\sim10^{7}$ transient sources per night in the $ugrizy$ filters. \citet{johnson03-19} predict that $\sim18\%$ of Galactic LMXBs will have their orbital periods (in the range of 10 min to 50 days) determined from LSST variability data. Difference imaging using LSST data will also help to identify counterparts to transient X-ray sources \citep[e.g., previous work in M31;][]{williams07-04, barnard09-12, barnard06-15}, which can then be used to estimate orbital periods from the optical/X-ray flux ratio \citep{revnivtsev04-12}. High-cadence optical observations are thus a powerful tool that can be used to help determine compact object types. At radio wavelengths, next-generation radio telescopes having an order of magnitude increase in sensitivity will make nearby galaxy monitoring campaigns feasible, and at the very least offer precise astrometric follow-up for luminous XRBs/outbursts. 
However, all multiwavelength approaches require X-ray detection of sources to identify XRB candidates.

Lastly, future gravitational wave detections in well-studied nearby galaxies will advance our understanding of progenitor populations and environments conducive to these events.
For example, the NS-NS merger GW170817 observed by LIGO/VIRGO was also detected as a short gamma-ray burst by \fermi\ \citep{abbott10-172, abbott01-19} and at X-ray wavelengths by \chandra\ 9 days after merger \citep{haggard10-17, troja11-17}, corresponding to a likely off-axis orientation.

\section{Experimental Requirement Necessary to Answer Key Question}

A combination of X-ray observatories with unique instrumental specifications are required to address the fundamental questions outlined here. For deep surveys of, e.g., nearby galaxies and the Galactic Center, a combination of $\sim1$ ms time resolution to detect pulsations, a large field of view ($\sim0.5$ deg$^{2}$) and collecting area ($>1$ m$^{2}$ at 1 keV) for efficient surveys, and moderate to exquisite angular resolution ($\lesssim5$\arcsec\ PSF HEW to resolve nearby point sources in crowded fields) will be required to transform our understanding of XRB populations at $\sim0.1-10$ keV energies. 

Finally, we should also point out that sub-luminous outbursts from BH and NS populations, such as very fast X-ray transients \citep[VFXTs, e.g.,][]{degenaar02-09, degenaar09-15} and supergiant fast X-ray transients \citep[SFXTs, e.g.,][]{sguera12-05, sguera07-06} are important in addressing our science goals. 
This motivates the need for a transient X-ray monitor having a large field of view ($>5000$ deg$^{2}$), rapid response time ($<1$ hr) for follow-up, sensitivity of $\sim10^{-8}$ erg cm$^{-2}$ s$^{-1}$ in 1 s to detect bursts in the Milky Way and nearby galaxies, angular resolution of $\sim1$\arcmin\ to aid with localization, optimized time resolution and collecting area for timing studies and to mitigate pile-up for bright Galactic source populations, and hard X-ray capability in the $\sim1-100$ keV energy range to probe obscured sources.
These specifications will be especially useful for time domain studies of transients given the wealth of all-sky multiwavelength data in the coming decade.

\newpage

\microtypesetup{protrusion=false}

\bibliographystyle{likeapj}

\end{document}

%% file: definitions_2017.tex
\newcommand{\be}  {{B\emph{e}}}

\newcommand{\msun}{${M}_{\odot}$}

\newcommand{\simgt}{\lower 2pt \hbox{$\, \buildrel {\scriptstyle >}\over {\scriptstyle\sim}\,$}}
\newcommand{\simlt}{\lower 2pt \hbox{$\, \buildrel {\scriptstyle <}\over {\scriptstyle\sim}\,$}}

\newcommand{\xmmn}{{\emph{XMM-Newton}}}
\newcommand{\chandra}{{\emph{Chandra}}}

\newcommand{\nustar}{{\emph{NuSTAR}}}

\newcommand{\fermi}{{\emph{Fermi}}}

\newcommand{\apj}{{\it ApJ~}}
\newcommand{\apjs}{{\it ApJS~}}
\newcommand{\apjl}{{\it ApJL~}}

\newcommand{\aap}{{\it A\&A~}}

\newcommand{\aapr}{{\it A.\&ARv.~}}

\newcommand{\mnras}{{\it MNRAS~}}

\newcommand{\nat}{{\it Nature~}}

\def\es{erg~s$^{-1}$}

\def\msun{$M_{\odot}$}

\def\logxr1{$\log (f_X/f_R) \le -1$}

\def\lx{$L_{\rm{X}}$}

\def\lxol2{$\log (f_X/f_O) < -2$}

\def\ran{$-2 < \log (f_X/f_O) < -1$}
\def\ran1{$-1 < \log (f_X/f_O) < +1$}

\def\chandra{{\it Chandra}}

\def\2df{{\it 2dfGRS}}

\def\exi{\begin{equation}}
\def\exo{\end{equation}}

\def\spose#1{\hbox to 0pt{#1\hss}} 
\def\approxlt{\mathrel{\spose{\lower 3pt\hbox{$\sim$}}
        \raise 2.0pt\hbox{$<$}}}
\def\approxgt{\mathrel{\spose{\lower 3pt\hbox{$\sim$}}
        \raise 2.0pt\hbox{$>$}}}

\renewcommand{\msun}{\hbox{${M}_{\odot}$}}

\renewcommand{\simgt}{\lower 2pt \hbox{$\, \buildrel {\scriptstyle >}\over {\scriptstyle\sim}\,$}}
\renewcommand{\simlt}{\lower 2pt \hbox{$\, \buildrel {\scriptstyle <}\over {\scriptstyle\sim}\,$}}

\newcommand{\arcsec}{\mbox{$^{\prime\prime}$}}
\newcommand{\arcmin}{\mbox{$^{\prime}$}}

\renewcommand{\chandra}{{\emph{Chandra}}}
\newcommand{\swift}{{\emph{Swift}}}

\def\apj{ApJ}                 
\def\apjl{ApJ}                
\def\apjs{ApJS}               
\def\aap{A\&A}                
\def\aapr{A\&A~Rev.}          
\def\mnras{MNRAS}             
\def\nat{Nature}              
